# The anomaly Cu doping effects on LiFeAs superconductors

L.Y. Xing, H. Miao, X.C. Wang*, J. Ma, Q.Q. Liu, Z. Deng, H. Ding, C.Q. Jin*

*Beijing National Laboratory for Condensed Matter Physics, Institute of Physics, Chinese Academy of Sciences, Beijing 100190, China*

## Abstract

The Cu substitution effect on the superconductivity of LiFeAs has been studied in comparison with Co/Ni substitution. It is found that the shrinking rate of the lattice parameter *c* for Cu substitution is much smaller than that of Co/Ni substitution. This is in conjugation with the observation of ARPES that shows almost the same electron and hole Fermi surfaces (FSs) size for undoped and Cu substituted LiFeAs sample except for a very small hole band sinking below Fermi level with doping, indicating little doping effect at Fermi surface by Cu substitution, in sharp contrast to the much effective carrier doping effect by Ni or Co.

**PACS numbers**: 74.70.Xa，74.25.-q，74.62.-c

**Introduction**

Since the discovery of La[O$_{1-x}$F$_x$]FeAs in 2008 [1], various classes of iron based superconductors such as "122" [2], "111" [3] or "11" [4] are reported [5-7]. These iron based superconductors contain superconducting [FePn(Se)] (where Pn is pnictide element of As or P) layers which are interlaced by charge carrier reservoir layers. Most of their parent compounds are in the form of antiferromagnetic spin density waves (SDW) states. The antiferromagnetism can be suppressed by either introducing charge carrier or applying pressure, leading to superconductivity. Superconductivity can be induced by various ways of element substitution either in the plane of [FePn(Se)] layer or out of the plane. For instance, in the case of BaFe$_2$As$_2$, K-substitution at the Ba site, Co or Ni substitution at the Fe site, or P substitution at the As site will induce superconductivity respectively [2, 8-10]. The partial replacement of Ba$^{2+}$ ion by K$^+$ ion will introduce hole-like charge carrier to the system while chemical pressure is applied when As atoms are substituted partially by P atoms. However, the effect of substitution at Fe site is quite different [11~21]. It seems that both Co or Ni substitution at Fe site introduce itinerant electrons as experimentally indicated by ARPES [15-17] & transport measurements [16-20] or X-ray emission spectroscopy measurements [21,22]. On the other hand the density-functional studies of the Fe$_{1-x}$Cu$_x$Se show that although Cu serves as an effective electron dopant, it is still a source of strong scattering[23]. Recently, the ARPES studies of Ba(Fe$_{1-x}$Cu$_x$)$_2$As$_2$ [24] and the work on NaFe$_{1-x}$Cu$_x$As[25] have revealed that part of electrons doped by substitution of Cu are almost localized.

To date, the research on this issue is mainly limited to the Ba-122 system since it is relative easy to grow high quality single crystals. In fact "111" type iron based superconductors are unique. In the structure of "111"type compounds, the [FePn] layers are intercalated with two layers of alkali metals atoms [3, 26]. The "111" system shows systematic evolution of superconductivity as function of Co/Ni doping [27] or pressure [28-30]. The crystal can be easily cleaved and results in an equivalent and neutral counterparts with identical surface versus bulk electronic structures, which is favored by ARPES [31-33]. Moreover, it can be referred as an electron over doped superconductor [34, 35]. When doped with Co/Ni in the parent LiFeAs, it presents no SDW transition and no dome like superconducting phase diagram but a linear suppression of $T_c$ [27], making this compound a good candidate for studying the effect of TM substitution. So far, the reported chemical doping effect of LiFeAs is limited to Co/Ni substitution [27, 33]. In this work, we study the effect of Cu substitution in comparison with Co/Ni substitution on the superconductivity of single crystals in LiFeAs system. We found that the behavior of Cu substitution is quite different from that of Co/Ni substitution in that most of the 3d electrons from Cu dopant are mostly localized.

I. **Experimental details**

Single crystals of LiFe$_{1-x}$TM$_x$As(TM= Cu, Co/Ni) were grown by self-flux method, using Li$_3$As, As, and Fe$_{1-x}$TM$_x$As powder as the starting materials. The precursor Li$_3$As was obtained by mixing Li lump and As powder, which was then sealed in an evacuated titanium tube and sintered at 650℃ for 10h. Fe$_{1-x}$TM$_x$As were prepared by mixing Fe, Cu (or Co/Ni) and As powder thoroughly, pressed into pellets, sealed in a evacuated quartz tube, and sintered at 700℃ for 30h. To ensure the homogeneity of the product, these pellets were grounded and heated again. The stoichiometric amount of Li$_3$As, Fe$_{1-x}$TM$_x$As and As powder were weighed according to the element ratio of Li(Fe$_{1-x}$TM$_x$)$_{0.3}$As. The mixture was grounded and put into alumina crucible and sealed in Nb crucible under 1 atm of Argon gas. The Nb crucible was then sealed in the evacuated quartz tube and heated to 1100℃ and cooled slowly down to 700℃ at a rate of 3℃/hr to grow single crystals. The obtained LiFe$_{1-x}$TM$_x$As single crystals have the typical size of 10mm * 6mm * 0.5mm, as shown in Fig. 1a. All sample preparations, except for sealing, were carried out in the glove box filled with high purity Argon gas.

The element composition of the LiFe$_{1-x}$TM$_x$As single crystals was checked by energy dispersive x-ray spectroscopy (EDS). These single crystals were characterized by x-ray diffraction. The transport measurements were carried on commercial physical properties measurement system (PPMS) using the four probe method. The dc magnetic susceptibility was measured with a magnetic field of 30 Oe using a superconducting quantum interference device (SQUID). ARPES studies were performed at beam lines PGM and Apple-PGM of the Synchrotron Radiation Center, Wisconsin, equipped with Scienta R4000 analyzer and SES 200 analyzer, respectively. The energy and angular resolutions of the ARPES measurements were set at 20-25 meV and 0.2°, respectively. The samples were cleaved in situ and measured at 30 K under a vacuum of $5 \times 10^{-11}$ torr. The incident photon energy was chosen to be hν = 51 eV.

## II. Results & discussions

The element composition checked by EDS is close to the nominal one. Thus, here the nominal concentration is used in the sample chemical formula. The typical x-ray diffraction pattern of the *00l* reflections for LiFe$_{1-x}$TM$_x$As single crystal is shown in Fig. 1a. From the diffraction pattern, the lattice constant *c* was calculated and the obtained c-axis values are plotted as a function of doping level *x* for Cu, Co/Ni substituted samples as shown in Fig. 1b, indicating a successful chemical substitution. However, in the case of Cu substituted samples, the lattice constant *c* shrinks by ~0.06% at the doping level *x* = 0.06; while for the Co/Ni substituted samples at the same doping level the *c* value decreases by ~0.3% that is much larger than that of Cu substituted samples. This will be further discussed in conjugation with ARPES measurements.

Fig. 2 presents the transport and magnetic data of Cu substituted LiFeAs single crystals. The temperature dependence of in-plane resistivity ρ is shown in Fig. 2a and the magnetic susceptibility in both zero field cooling (ZFC) and field cooling (FC) modes are shown in Fig. 2b. For undoped LiFeAs crystal, the resistivity drops sharply to zero at ~17K with a narrow superconducting transition width ⊿T ~1.1K and the residual resistivity ration (RRR), defined as the ratio of the resistivity at 300K and residual resistivity $\rho_0$ which is determined by extending from the range right above Tc is found to be 60. Upon Cu doping, the effect of its scattering on electron mobility increases and lead to the increase of resistivity as shown in Fig. 2a, implying the localization of doping carriers from Cu that plays more like an impurity center. The magnetic susceptibility of LiFeAs crystal shown in Fig.2b suggests bulk superconductivity with $T_c$ ~16K which is defined by the bifurcation point between ZFC and FC magnetic susceptibility in consistent with transport data. As shown in Fig. 2a and Fig. 2b the $T_c$ of Li(Fe$_{1-x}$Cu$_x$)As is gradually suppressed reaching to ~3K at the Cu doping level of 7%. The $T_c$ extracted from the resistivity and magnetic measurements as a function of doping level are plotted in Fig. 2c, showing an almost linear dependence on Cu doping level. The red line represents a linear fitting to $T_c$ change as function doping level, which demonstrates $T_c$ decreases at a rate about 1.9K per 1% Cu dopant in Li(Fe$_{1-x}$Cu$_x$)As.

For Co substituted LiFeAs single crystal samples, the ρ-T curve and magnetic susceptibility measurements are shown in Fig. 3a and Fig. 3b, respectively. All the samples show a sharp superconducting transition. The $T_c$ decreases with increasing Co doping level and is suppressed down to ~4K by 12% Co doping. Fig. 3c presents the linear fitting result for the data of $T_c$ versus Co doping level showing an approximately 1K suppression rate per 1%

Co doping. Fig. 4a and Fig. 4b are the transport and magnetic properties for LiFe$_{1-x}$Ni$_x$As crystals respectively, while the $T_c$ value versus Ni-doping level is plotted and fitted linearly as shown in Fig. 4c. $T_c$ is linearly suppressed by Ni doping with a rate of about 2.2K per 1% Ni doping.

We found that the suppression rate of $T_c$ for Ni substitution is twice of that for Co substitution implying that Co & Ni substitution introduce one & 2 more itinerant electrons, respectively, which is consistent with the change of lattice parameter. The similar behavior had been observed in Co and Ni doped Ba 122 system[36].

To further verify the localization tendency of Cu doped electron which is different with the Co/Ni substitution with less change of lattice parameter for Cu doping as shown in Fig.1(b), we measured the electronic structure by using APRES technique in order to get a straightforward electronic structure picture of Cu doped LiFeAs. It has been experimentally [31] proved that the surface of LiFeAs preserves its bulk properties. Therefore ARPES reflects the intrinsic properties for the LiFeAs crystals. Previously ARPES are used to study the Fermi surface evaluation when Fe is partially substituted by Co [33]. The results show that Co substitution introduces electron type charge carriers and results in chemical potential shifting upwards, indicating electron doping. The Fermi surface (FS) mappings along the Γ-M high symmetry line from ARPES for undoped and Cu 6% doped LiFeAs are shown in Fig. 5a and Fig. 5b, respectively. To check how FSs change with Cu substitution, extracted $k_F$ locus were plotted in Fig. 5a, 5b and 5e, respectively. Red circles and green triangles represent undoped and Cu 6% doped Li(Fe$_{1-x}$Cu$_x$)As, respectively. The results reveal no significant difference in these two different crystals, except for the small hole FS observed in LiFeAs which disappeared in the Cu 6% doped Li(Fe$_{1-x}$Cu$_x$)As, indicating small electron doping effect. Additionally, the normal state (T = 30 K) high resolution ARPES intensity plots along the Γ-M high symmetry line are shown in Fig. 5c and Fig. 5d, respectively. The incident light was set to 51 eV ($k_z = 0$) with its polarization perpendicular to the mirror plane to select odd symmetry orbitals. Red and green solid circles are used to extract the band dispersion, which generates the small hole FS in LiFeAs. It is clear that the extracted band is crossing $E_F$ in LiFeAs but sinking below $E_F$ in Cu 6% sample. Our results show that, unlike Co and Ni substituted LiFeAs [33], where extra electrons cause Fermi level shifted, the FSs of Cu substituted LiFeAs remained almost intact. Hence the 3d electrons from Cu dopant in LiFeAs are more localized and contribute little to the FSs. T. Berlijn et.al. studied disorder effects of Co and Zn substitution in Ba122 system [37]. They found that the calculated Femi surface behaviors of Zn substitution induced deep impurity level that is quite different from the effect of Co doping. Recently S. Ideta et.al. reported, strong localization effects in Zn substituted BaFe$_2$As$_2$, wherein all the extra electrons are localized at the state of ~10 eV below FSs and do not contribute to the chemical potential shift at all [38]. Our results can be explained from the d-band partial density of states of Co, Ni and Cu in iron based superconductors. The behavior of Co or Ni substitution follows rigid band model due to the fact that the d-bands of Co and Ni overlap with Fe d-band and are featureless compared with Fe d-band; whereas Cu presents deeper impurity potential of ~4 eV [22], which localizes most of the Cu 3d orbital electrons. This is in consistent with the less contraction of lattice parameters by Cu substitution compared with Co/Ni substitution wherein itinerant electrons are induced. Although there is a small electron doping effect by Cu substitution in LiFeAs system, the doped coherent electrons would be even smaller considering the effect of disorder with strong impurity potential [38], hence the suppression of $T_c$ by Cu substitution is mainly from strong impurity scattering instead of carrier density change. Here we found that Cu doping results in the sinking of the small hole band below $E_F$ in LiFe$_{0.94}$Cu$_{0.06}$As that contributes to approximately only 1% mobile electron carriers(normalized to 0.17 electron / Cu doping). Therefore the Cu doping is much localized in LiFeAs system, quantitatively different from Cu doped other systems[23][24][25]. Taking an example, the practical doped mobile carriers in Ba(Fe$_{1-x}$Cu$_x$)$_2$As$_2$[24] are about 1 electron per Cu as calculated from the change of the Fermi Surface volume, much higher than 0.17 electron per Cu in our LiFeAs system.

On the other hand, it is interesting to compare Cu doping with Ru doping. Since Ru is isovalent to Fe (within the same column in periodic table), Ru doping will not introduce carriers theoretically. The experiments of ARPES verified that Ru is isoelectronic substitution [39]. Therefore Cu doping versus Ru doping represent two typical cases that do not introduce carriers. But the mechanism is different, for Cu doping case the "carriers" are localized while for Ru doping there is no additional carriers induced.

In summary, series of LiFe$_{1-x}$Cu$_x$As single crystals were grown by self-flux method. Based on systematic investigations of superconducting transitions, crystal versus electronic structure evolution with Cu doping level, we found that behaviors of Cu substitution are different from those of Co/Ni substitution in both change rates of Tc as well as lattice parameters as function of doping level. ARPES measurements indicated that most of the 3d valence electrons from Cu dopant are localized, resulting in an almost intact Fermi surfaces for Cu doped LiFeAs except for a very small hole band sinking below Fermi level which is also quite different from the doping effects in other systems.

**Acknowledgements**

The work is supported by NSF & MOST of China through research projects. We are grateful to PGM & Apple PGM of the Synchrotron Radiation Center Wisconsin for supporting ARPES experiments.

**Figure Captions:**

Figure 1. (a) The typical XRD patterns for LiFe$_{1-x}$TM$_x$As single crystals. The inset is the photo of LiFeAs single crystal with typical size of 10mm * 6mm * 0.5mm; (b) The dependence of lattice constant c on doping level $x$ for LiFe$_{1-x}$Tm$_x$As single crystal (Tm = Co, Ni and Cu).

Figure 2. (a) The temperature dependence of resistivity for LiFe$_{1-x}$Cu$_x$As single crystal. (b) The magnetic susceptibility of LiFe$_{1-x}$Cu$_x$As single crystal. (c) The critical temperature plotted as a function of Cu-doped level $x$. The red line is the linear fit of $T_c$ versus Cu-doped level $x$ for both the magnetic susceptibility measurement data and resistance measurement data.

Figure 3. (a) The temperature dependence of resistivity for LiFe$_{1-x}$Co$_x$As single crystal. (b) The magnetic susceptibility of LiFe$_{1-x}$Co$_x$As single crystal. (c) The critical temperature plotted as a function of Co-doped level $x$. The red line is the linear fit of $T_c$ versus Co-doped level $x$ for both the magnetic susceptibility measurement data and resistance measurement data.

Figure 4. (a) The temperature dependence of resistivity for LiFe$_{1-x}$Ni$_x$As single crystal. (b) The magnetic susceptibility of LiFe$_{1-x}$Ni$_x$As single crystal. (c) The critical temperature plotted as a function of Ni-doped level $x$. The red line is the linear fit of $T_c$ versus Ni-doped level $x$ for both the magnetic susceptibility measurement data and resistance measurement data.

Figure 5. (a), (b) ARPES intensity at $E_F$ of LiFeAs and Cu 6% doped LiFeAs with photon energy at 51eV. The intensity is obtained by integrating the spectra within ±10 meV with respect to $E_F$. (c), (d) ARPES high resolution cut along high symmetry line Γ-M at $k_z$ = 0. (e) Extracted $k_F$ locus of LiFeAs and Cu 6% doped LiFeAs.

Figure 1 (a, b)

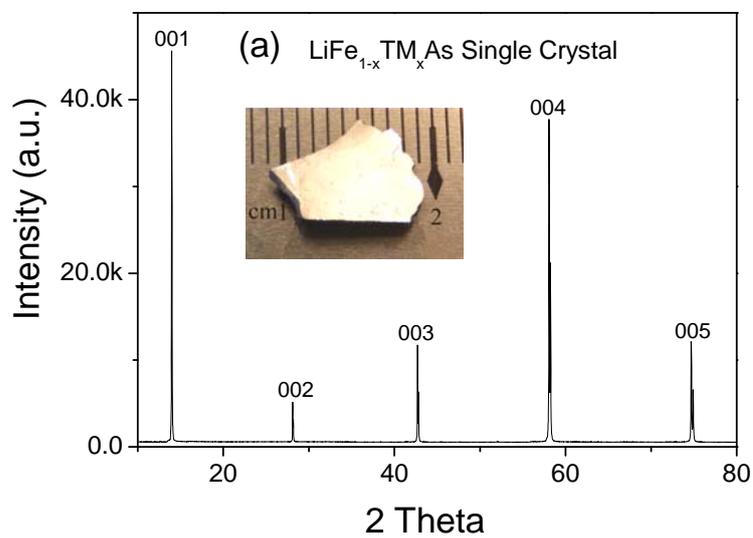

Figure 1 (a)

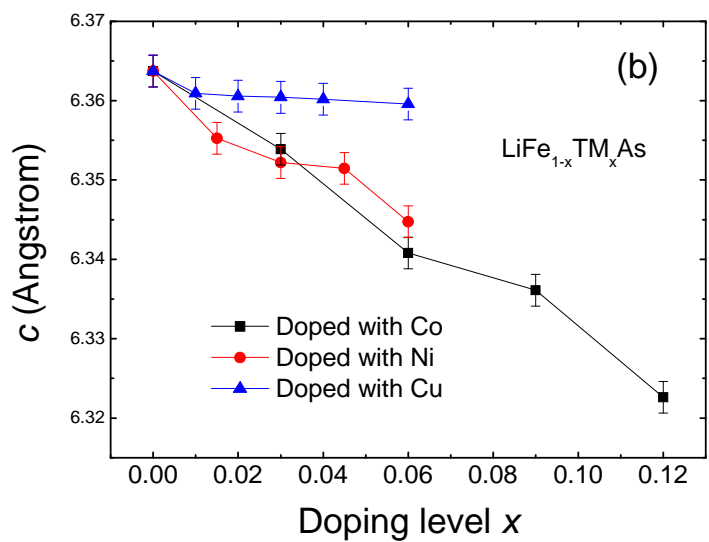

Figure 1 (b)

Figure 2 (a, b, c)

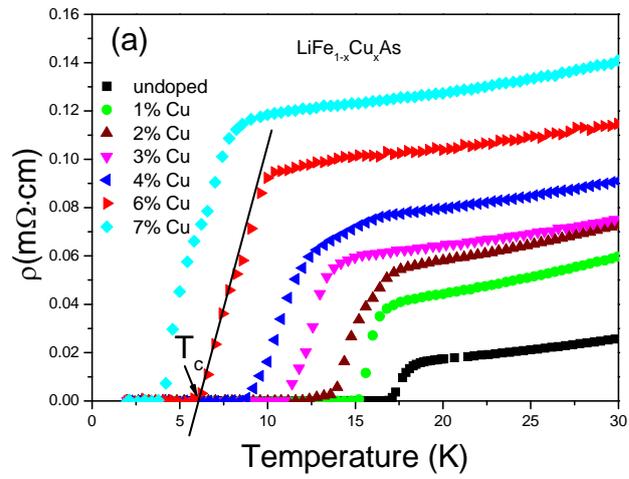

Figure 2 (a)

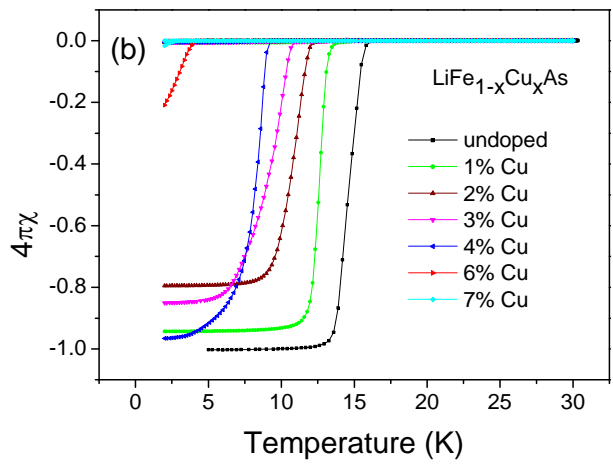

Figure 2 (b)

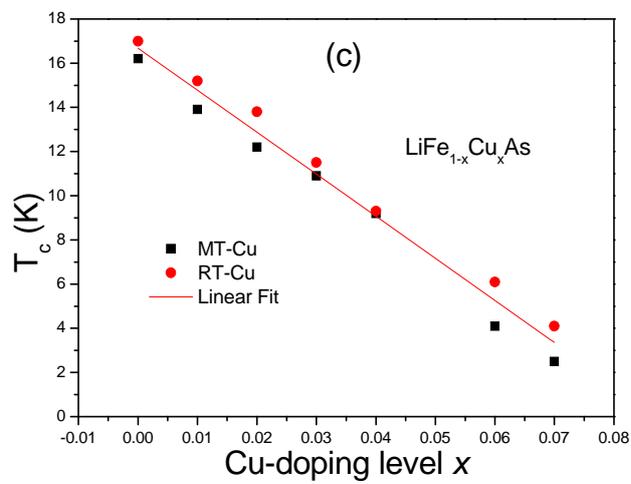

Figure 2 (c)

Figure 3 (a, b, c)

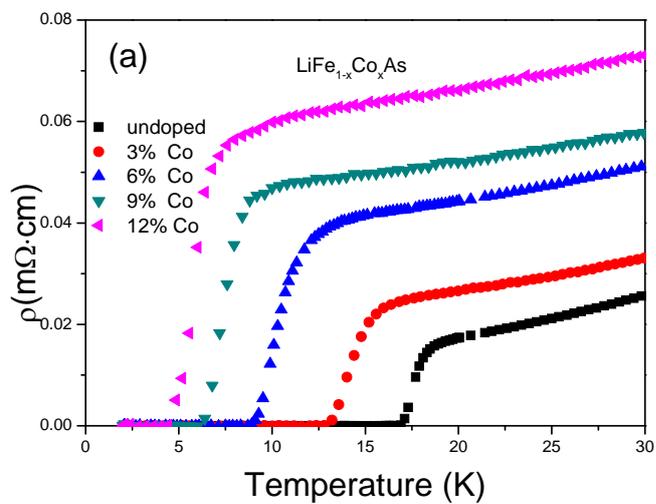

Figure 3 (a)

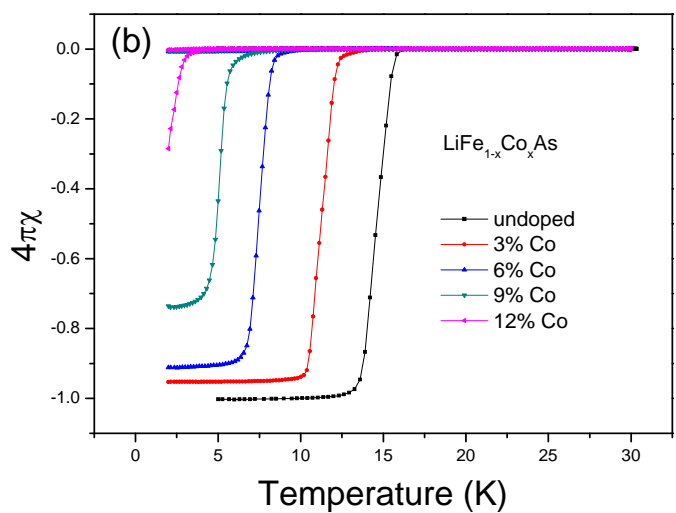

Figure 3 (b)

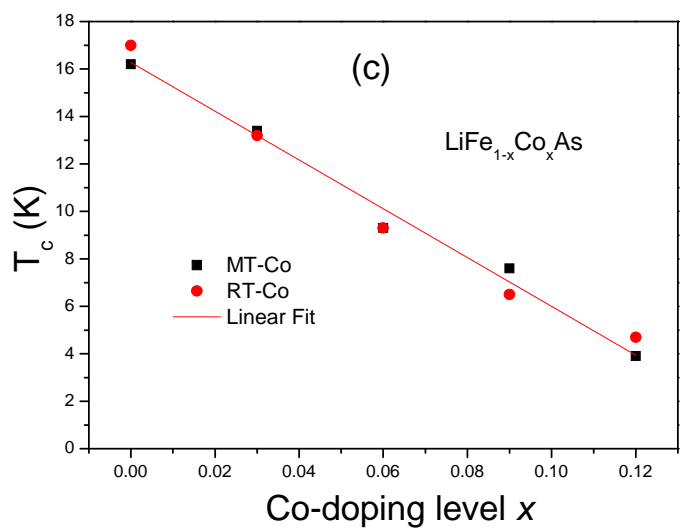

Figure 3 (c)

Figure 4 (a, b, c)

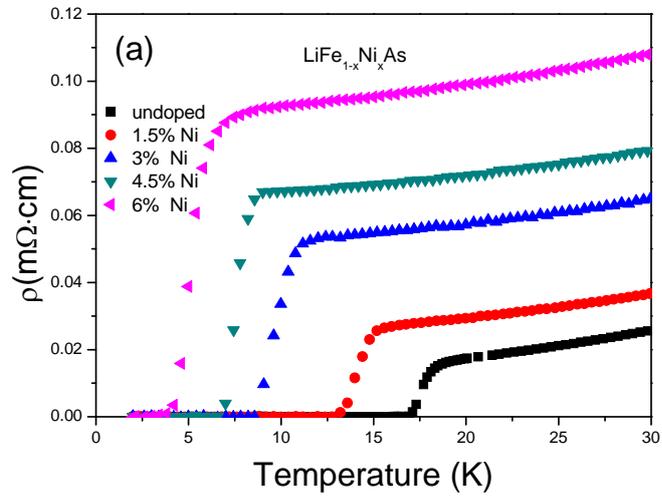

Figure 4 (a)

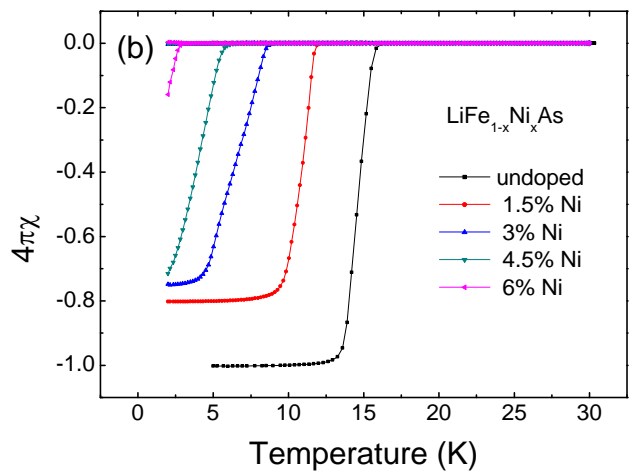

Figure 4 (b)

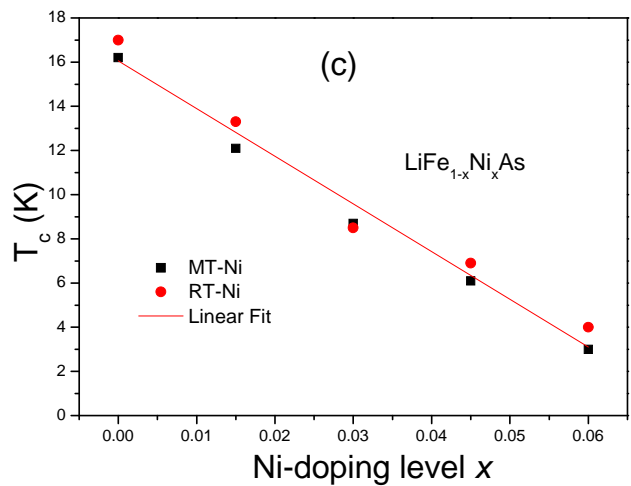

Figure 4 (c)

Fig. 5 (a, b, c, d, e)

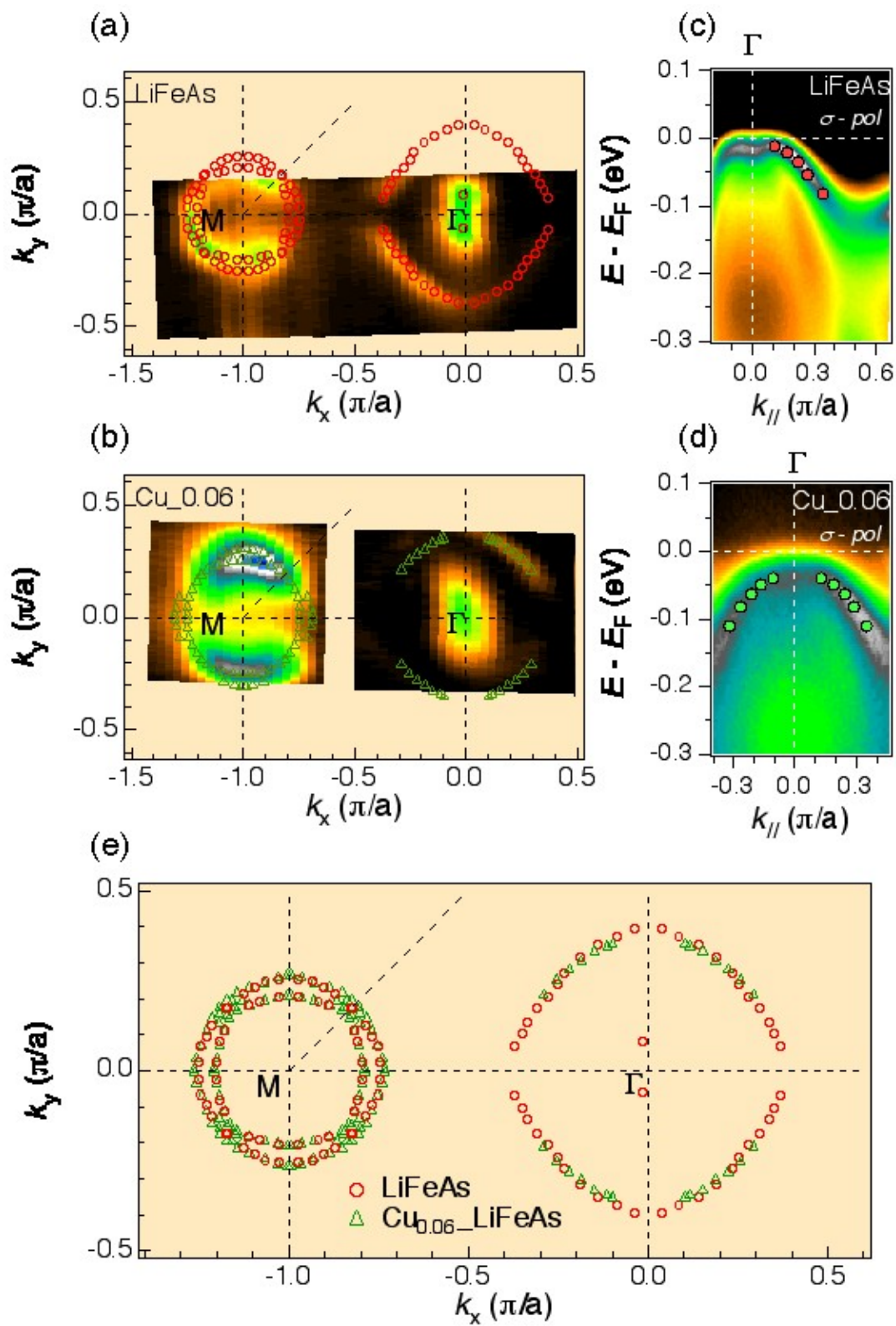